# Ballistic bit addressing in a magnetic memory cell array

*H.W. Schumacher*

Physikalisch-Technische Bundesanstalt, Bundesallee 100, D-38116 Braunschweig, Germany.

E-mail: hans.w.schumacher@ptb.de,
phone: +49 (0)531 592 2414,
fax: +49 (0)531 592 2205

## Abstract

A ringing free bit addressing scheme for magnetic memories like MRAM (magnetic random access memory) is proposed. As in standard MRAM addressing schemes the switching of a selected cell is obtained by the combination of two half-select field pulses. Numerical solutions of a single spin model of an MRAM cell show that the pulse parameters can be chosen such that the application of the half select pulse induces a full precessional turn of the magnetization (no switch) whereas the superposition of two half select pulses induces a half precessional turn (switch). With well adapted pulse parameters both full-select *and* half-select switching occurs on ballistic trajectories characterized by the absence of ringing after magnetic pulse decay. Such ballistic bit addressing allows ultra high MRAM clock rates.

# Article

The fastest way to reverse the magnetization of a magnetic memory cell is the so-called ballistic magnetization reversal (1,2). This switching mechanism uses a fast-rising and ultra-short pulse which is oriented mainly perpendicular to the equilibrium magnetization to induce a 180° precessional turn of the magnetization. By properly adapting the field pulse parameters to the half of the precession period (i.e. to a 180° turn) an optimal "ballistic" trajectory can be obtained. In this ballistic case the magnetization turns directly from the initial to the final orientation without any remaining magnetic precession ("ringing") after field pulse decay (3,4,5,6). For fast operation of a magnetic memory such as an MRAM (magnetic random access memory) (7,8) such ringing must be avoided as the damping time can take several nanoseconds which would eventually limit the clock speed of the memory device. For practical MRAM operation, however, a principal problem arises. The standard MRAM architecture consists of two sets of write lines running row- and column wise and a memory cell at each intersection of the rows and columns. Bit addressed switching (i.e. selective switching of only one cell of the array of cells) is performed by applying so-called half select field pulses to one row and one column at the same time. One half-select pulse alone is not sufficient for bit addressing i.e. to switch the magnetization of a cell whereas the superposition of the two half-select pulses at the intersection is. It has been verified experimentally that also in this case ultra fast quasi ballistic switching is possible (9,10). But what about the other cells subjected to the half-select pulse only? Here, a strong ringing can still occur which will limit the MRAM clock speed.

In this work a way is proposed to solve the problem of ringing in bit addressed MRAM switching. By numerical calculation of the magnetization dynamics of the cell in a simple single spin model it is shown that one can find pulse parameters leading to ringing free ballistic trajectories both for the case of switching (full bit selection) and for the case of non switching (half-selection). This ballistic bit addressing scheme allows the realization of ultimate MRAM clock speeds.

The MRAM bit structure studied in this work is shown in Fig. 1. The magnetic cell is a rotational ellipsoid of permalloy (NiFe) having a saturation magnetization of $4\pi M_S = 10800$ Oe and demagnetizing factors of $N_X/4\pi = 0.00615$ (easy axis), $N_Y/4\pi = 0.01746$ (in-plane hard axis), and $N_Z/4\pi = 0.9764$ (out of plane) corresponding to ellipsoid dimensions of 500 nm x 200 nm x 5 nm (11). Other anisotropies than shape anisotropy are not taken into account. Before pulse application the magnetization $M$ is oriented in the –x direction. The half select and full select magnetic field pulses ($H_{HS}$, and $H_{FS}$, respectively) are always oriented along the intermediate axis of the ellipsoid (y-direction, corresponding to the in-plane hard axis of the memory cell). The amplitude and orientation of the two half select pulses running through the word and bit line are the same and therefore $H_{FS} = 2 \cdot H_{HS}$. Such field pulses could be realized by a pulse line design where the bit line (BL) and the word line (WL) run parallel in the vicinity of the cell as sketched in Fig. 1.

The time evolution of the magnetization response to the transverse field pulses is derived by numerically solving the Landau-Lifshitz-Gilbert (LLG) equation (12)

$$\frac{dM}{dt} = -\gamma(M \times H_{eff}) + \frac{\alpha}{M_S}\left(M \times \frac{dM}{dt}\right)$$

in the macro spin approximation where a homogeneous magnetization in the cell is assumed (1). In the LLG equation $\gamma$ is the gyromagnetic ratio, and $\alpha$, is the Gilbert damping parameter.

As already mentioned the aim is to find ringing free trajectories of the magnetization both for the application of a half-select (non-reversal) pulse and for the application of a full-select (reversal) pulse. This means that upon pulse termination *M* must be oriented along the –x (not reversed) direction for the half select case and along the +x (reversed) direction for the full select case. In these two cases *M* is already in equilibrium when the field pulse terminates and no magnetic ringing will occur afterwards. For the full select case (ballistic switching) this is generally done by applying a pulse of sufficient field amplitude $H_{FS}$ where the pulse duration $T_{Pulse}$ is matched to the duration of a half of the precession period (1,2) i.e. $T_{Pulse} = ½·T_{Prec}(H_{FS})$. Also for half-selection such a no-ringing criterion can be established. During application of the half select field pulse $H_{HS}$ the magnetization has to undergo exactly a *full* precessional turn and thus $T_{Pulse} = T_{Prec}(H_{HS})$ (13). If the two half select pulses are of the same amplitude $H_{HS}$ and duration $T_{Pulse}$ we find the criterion for ballistic bit addressing for our geometry: $T_{Pulse} = T_{Prec}(H_{HS}) = ½·T_{Prec}(2·H_{HS})$. (14)

In Fig. 2 the precession frequency in our model system is plotted as a function of the hard axis pulse field. $T_{Prec}(H)$ is derived from the numerical simulation of the response of *M* to a transverse step excitation of variable amplitude H. In the calculations an ideal pulse with vanishing rise time and a damping free magnetization ($\alpha = 0$) is considered. As seen in Fig. 2 for low transverse fields the precession period first weakly increases with H. Here, the shape anisotropy is dominating and M mainly precesses about internal (shape) anisotropy field. Conversely, for high fields the external field H is dominating and M precesses basically about the external field H. Here, $T_{prec}$ decreases with H as can be expected from simple ferromagnetic resonance considerations (15). The two regions are separated by a peak. Such a variation of the precession period with the in-plane hard axis field has also been observed experimentally in magnetic thin films (16). Note that on the low field side of the peak *M* does not overcome the hard axis during step excitation and thus no switching is possible. In Fig. 2 also ½·$T_{Prec}$(2·H) is displayed. The two curves intersect in two points. In both points of intersection the criterion $T_{Prec}(H)$ = ½·$T_{Prec}$(2·H) is fulfilled. However as the *M* has to overcome the hard axis during switching (which is not the case on the low field point) only the high field point of intersection (arrow) allows ballistic bit addressing. The ballistic bit addressing parameters are $T_{Pulse}$ = 344 ps, $H_{HS}$ = 36.75 Oe, and $H_{FS}$ = 2·$H_{HS}$ = 73.5 Oe. The two step response trajectories of $M = (m_X, m_Y, m_Z)$ *vs.* time for $H_{FS}$ and $H_{HS}$ are shown in Fig. 2 (b) and (c), respectively. As marked by the vertical arrow the duration of a half precession in (b) and the duration of a full precession are the same which should allow ballistic bit addressing.

The trajectories of ballistic bit addressing by adapted pulses for this set of parameters are shown in Fig. 3. In (a) the two pulses for half-select and full-select addressing are shown. Again, vanishing damping and vanishing rise and fall times of the pulses are assumed. The pulse duration is 344 ps and the amplitudes are 36.75 and 73.5 Oe for half and full selection, respectively. Fig. 3(b) shows the response to the full-select pulse $H_{HS}$. One can see that M performs exactly a half precessional turn about the pulse field during pulse application. After pulse decay the magnetization is reversed and no ringing is present. The response to the half-select pulse is shown in Fig. 3(c). Here, M performs a full precessional turn without overcoming the magnetic hard axis. Also here, M is in equilibrium upon pulse decay and a ringing free ballistic trajectory is observed. The pulse parameters would thus allow ballistic bit addressing and, as a consequence, ultimate MRAM clock rates.

Up to now only a very ideal system ($\alpha = 0$, vanishing rise and fall time) has been considered. Fig. 4 shows ballistic bit addressing trajectories for more realistic parameters. Here, the rise and fall times of the pulses are 100 ps and the Gilbert damping is $\alpha = 0.001$, a value which is feasible for high quality permalloy thin films. The demagnetizing factors and the saturation magnetization remain the same as given above. The two pulses are shown in Fig. 4(a). The change of damping and the rise and fall times lead to a shift of the optimum parameters for ballistic bit addressing. The pulse amplitudes are now $H_{HS} = 39$ Oe and $H_{FS} = 2 \cdot H_{HS} = 78$ Oe. The pulse duration is $T_{Pulse} = 345$ ps when measured at half maximum. The pulse starts to rise at 0 ps and fully decays to zero at a time of 395 ps. The switching trajectory of M generated by the full select pulse is shown in Fig. 4(b). Again M performs a half precessional turn, overcomes the hard axis, and switches to the reversed easy direction. Due to $\alpha \neq 0$ M is not exactly in equilibrium upon pulse decay and a slight ringing remains (arrow). However, the maximum angle mismatch out of the equilibrium direction is less then 1.5° and can be neglected. Also for the non-switching trajectory shown in Fig. 4(c) the ringing is practically fully suppressed. Upon pulse termination M has perfomed a full precessional turn and is not reversed. The angle mis-match of M from the equilibrium easy axis direction is even smaller than in the switching case and is hardly observable in Fig 4(c).

These calculations show that it is possible to obtain ballistic bit addressing in an MRAM array with practically ringing free trajectories both for half-select and full-select pulses. In the geometry discussed here the switching is obtained by hard-axis pulses only. The memory would thus be operated in the "toggle write" mode and M is switched back and forth with every full-select pulse application. For the set of parameters used in the simulation the bit addressing is completely terminated within less than 400 ps corresponding to a maximum MRAM clock rate of 2.5 GHz. These numerical values depend strongly on the device parameters (geometry and demagnetizing factors, further anisotropies, saturation magnetization, etc.). However, the order of magnitude is valid for a realistic memory cell. To give an example the increase of the saturation magnetization up to 13500 Oe (a value for a NiFe/CoFe free layer) results in shorter bit addressing times below 280 ps but with increased half-select pulse fields around 45 Oe. Note, that ballistic bit addressing should not only be fast but also stable. Upon pulse termination of the half- and full-select pulses the final magnetization is practically in equilibrium and thus near the lowest energy state. Thereby the probability of erroneous writing by half-select pulses e.g. due to thermal activation should be strongly reduced. Of course the simple model used here does not take into account any effects due to an inhomogeneous magnetization in the sample. Here, micromagnetic simulations and eventually an experimental test of ballistic bit addressing are necessary. Furthermore more complicated pulse geometries should be considered and tested for no-ringing trajectories. However, fast reversal experiments on microscopic memory cells show that macro spin simulations can well de-scribe the basic physics of ultra fast magnetization motion in MRAM cells (3,4,5). The road towards MRAM operating with GHz clock rates thus seems open.

**FIGURE CAPTIONS:**

**FIGURE 1:**

Geometry of the magnetic cell under consideration. The before pulse application the magnetization M is oriented along the –x direction (easy axis). The pulses are applied along the in-plane hard axis (y-direction). Two half select pulses $H_{HS}$ add up to a full select pulse $H_{FS}$. For such a field geometry the bit and the word line (BL, WL, respectively) run parallel along the easy axis of the cell.

**FIGURE 2:**

(a) Precession period during application of hard axis field steps *vs.* field amplitude H. $T_{Prec}(H)$ (black full line) peaks at the transition from anisotropy dominated trajectories to field dominated trajectories. $½·T_{Prec}(2H)$ is plotted in gray. At the high field intersection of the two plots (arrow) the conditions for ballistic bit addressing are fulfilled (H=36.75 Oe; 2H=73.5 Oe; $T_{Prec}$=344 ps). (b), (c) Components of $M = (m_X, m_Y, m_Z)$ *vs.* time at the field conditions for ballistic bit addressing. H=73.5 Oe (b), H=36.75 Oe (c). After 344 ps (arrow) M is reversed ($m_X$ =1) for the full select field in (b) and M is in its initial state ($m_X = 0$) for the half select field in (c).

**FIGURE 3:**

Ballistic bit addressing for vanishing damping ($\alpha=0$) and vanishing rise/fall time ($T_{rise} = T_{fall.} = 0$). (a) half select (gray) and full select (black) pulses. $H_{HS} = 36.75$ Oe, $H_{FS} = 73.5$ Oe, $T_{pulse} = 344$ ps. (b) Ballistic full select switching trajectory. M switches directly from the initial to the reversed easy axis orientation by a half precessional turn. No ringing occurs after pulse decay. (c) Ballistic half select switching trajectory. M undergoes a full precessional term and is aligned along the initial easy axis orientation upon pulse decay. Also after the half select pulse the ringing is suppressed.

**FIGURE 4:**

Ballistic bit addressing for weak damping and realistic pulse parameters; $\alpha = 0.001$ and $T_{rise} = T_{fall} = 100$ ps. (a) half select (gray) and full select (black) pulses. $H_{HS} = 39$ Oe, $H_{FS} = 78$ Oe, The pulse duration is 345 ps measured at half of the maximum amplitude. (b) Full select switching trajectory. M switches again by a half precessional turn. Only minor ringing occurs after pulse decay (arrow). (c) Half select switching trajectory. M undergoes a full precessional turn and is well aligned along the initial easy axis orientation upon pulse decay. The ringing is practically suppressed. The bit addressing is completed within 400 ps.

**References:**

(1)  M.  Bauer et al., Phys. Rev. B **61**, 3410 (2000).

(2)  J.  Miltat, G. Alburquerque, and A. Thiaville in *Spin Dynamics in Confined Magnetic Structures*, B. Hillebrands and K. Ounadjela (eds.) (Springer, Berlin, 2001).

(3) S. Kaka and S. E. Russek, Appl. Phys. Lett. **80**, 2958 (2002).

(4) H. W. Schumacher et al. Phys. Rev. Lett. **90**, 017201 (2003); H. W. Schumacher et al. Phys. Rev. Lett. **90**, 017204 (2003).

(5) Th. Gerrits et al., Nature **418**, 509 (2002).

(6) W. K. Hiebert, L. Lagae, and J. De Boeck, Phys. Rev. B **68**, 020402(R) (2003).

(7) S.S.P. Parkin et al., J. Appl. Phys. **85**, 5828 (1999),

(8) S. Tehrani et al., IEEE Trans. Mag. **36,** 2752 (2000).

(9) W. K. Hiebert, L. Lagae, J. Das, J. Bekaert, R. Wirix-Speetjens, and J. De Boeck, J. Appl. Phys. **93**, 6906 (2003).

(10) A. Krichevsky and M. R. Freeman, J. Appl. Phys. **95**, 6601 (2004).

(11) S. Chikazumi: The Physics  of Magnetism, p. 21-22, Wiley, New York, ISBN 0-88275-662-1.

(12) L. Landau, and E. Lifshitz, Phys. Z. Sowjetunion **8**, 153, (1935); T. L. Gilbert, Phys. Rev. **100**, 1243 (1955).

(13) H. W. Schumacher, C. Chappert, P. Crozat, R. C. Sousa, P. P. Freitas, M. Bauer, Appl. Phys. Lett. **80**, 3781 (2002).

(14) In principle also higher order switching and non-switching (4) could be considered. One could e.g. also find a half select field $H_{HS}$ where $T_{Prec}(H_{HS}) = 3/2 \cdot T_{Prec}(2 \cdot H_{HS})$ which would also allow ballistic bit addressing. In this work, however, only the simplest case of a full and a half precessional turn shall be discussed in detail.

(15) C. Kittel, *Introduction to Solid State Physics* (Wiley, New York, 1976), 5th ed.

(16) M. Buess, T.P.J. Knowles, U. Ramsperger, D. Pescia, C.H. Back, Phys. Rev. B **69**, 174422 (2004).

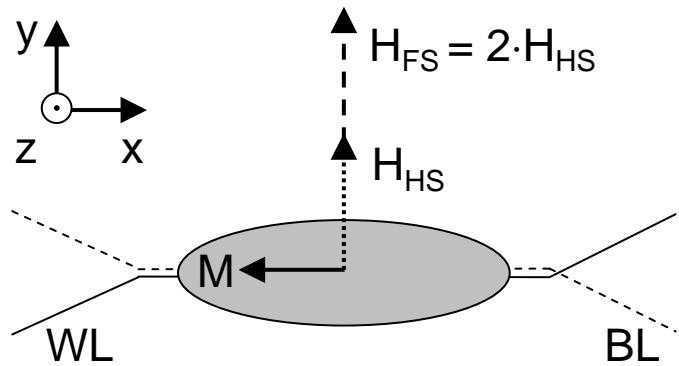

H. W. Schumacher
Figure 1

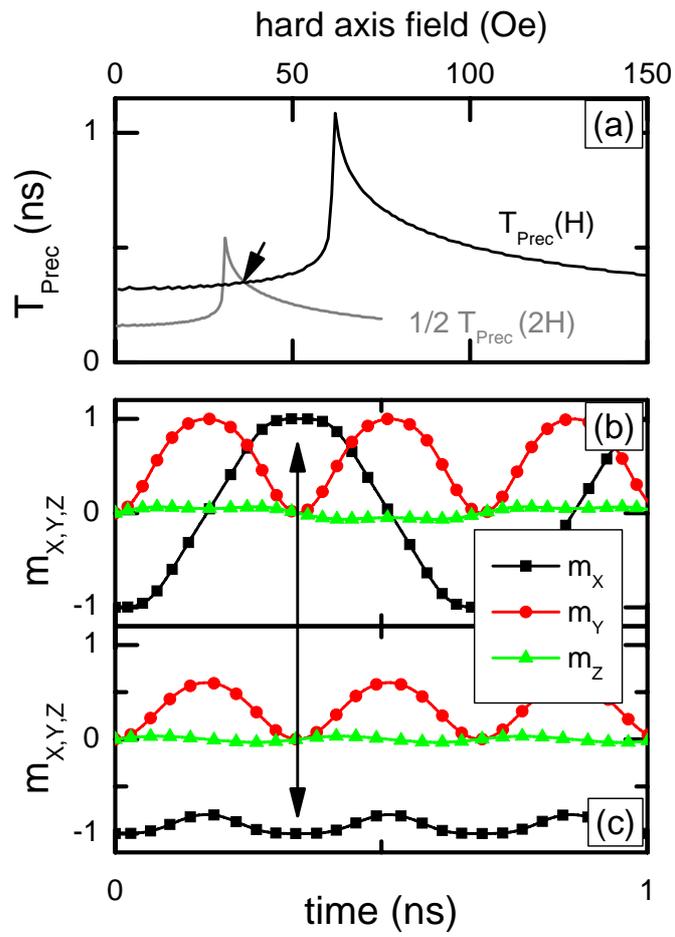

H.W.Schumacher
Figure 2

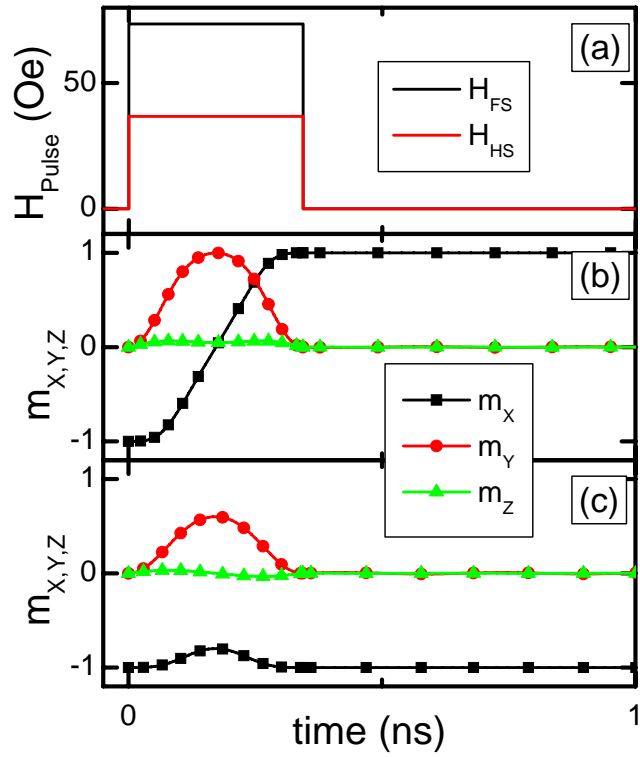

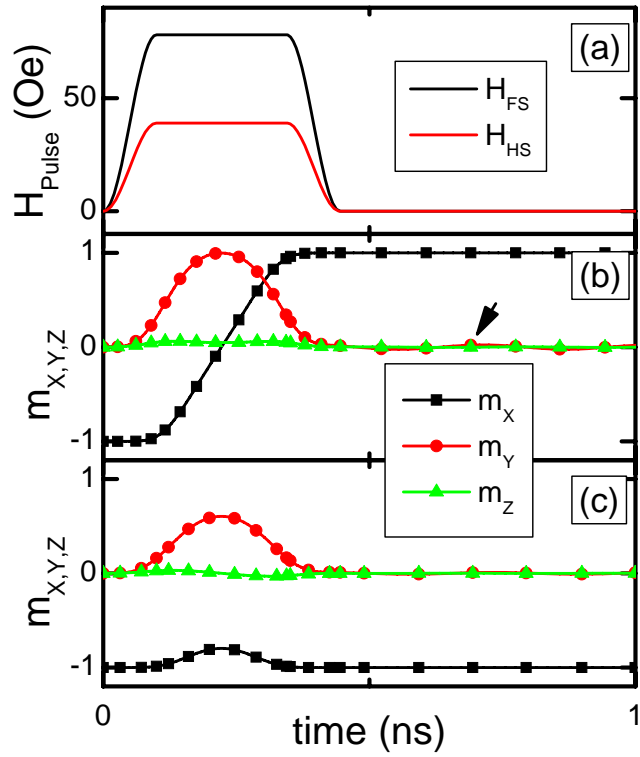

H.W.Schumacher
Figure 4